# Thermoelectric signature of the nematic phase iron-based superconductor

Marcin Matusiak[*], Michał Babij

*Institute of Low Temperature and Structure Research, Polish Academy of Sciences,*

*ul. Okolna 2, 50-422 Wroclaw, Poland*

**Studies of the copper-based superconductors demonstrate how their phase diagram becomes more complex as experimental probes improve, able to distinguish among subtly different electronic phases. One of those phases, nematicity, has become the matter of great interest also in the iron-based superconductors, where it is detected deep in the tetragonal state. Here we present the evolution of the in-plane Nernst effect anisotropy in the strain detwinned $Ca(Fe_{1-x}Co_x)_2As_2$ single crystals. We interpret the observed behaviour using an approach developed to describe the nematic order parameter in liquid crystals. We also apply the same model to data from other superconductors: $Ba(Fe_{1-x}Co_x)_2As_2$ and $YBa_2Cu_3O_y$. We conclude the observed broken rotational symmetry of the electronic system is a consequence of the emerging thermodynamic electronic nematic order at a temperature much higher than onset of the magnetic and structural transitions.**

**Introduction**

The phase diagram of the copper-based superconductors, which initially consisted of the superconducting and antiferromagnetic phases [1] was subsequently supplemented with regions of occurrence of other electronic orders. These include the pseudo-gap [2], spin density wave (SDW) [3 as well as three-dimensional (3D) and two-dimensional (2D) charge density wave (CDW) [4,5]. It appears the presence of some kind of electronic nematic order is a pervasive characteristic of unconventional superconductors [6,7], which applies also to iron-based superconductors, where the electronic nematic phase is a distinguished feature of their phase diagram [8]. While there is no agreement yet about a mechanism that leads to breaking of rotational symmetry in these electron systems, the phenomenon itself has gathered wide experimental support. It was extensively investigated by spectroscopic [9,10] and macroscopic [11,12] probes including studies of the thermoelectrical response anisotropy [13,14]. The latter has turned out to be an invaluable source of information about the nematic phase in cuprates [15,16]. Furthermore, it was argued that this method could detect a type of nematicity that was unreachable by electrical resistivity studies [15].

In this work we report measurements of the Nernst effect anisotropy, which indicate emergence of the nematic phase in the electron doped 122 iron-based superconductors. Notably, the data suggest that at high temperatures we do not observe fluctuations of the low-temperature nematic order, but rather a development of a fully set nematicity. Such a possibility was early discussed by M.A. Tanatar et al. [17], and similar conclusions were drawn on the basis of the magnetic torque [9] and point contact spectroscopy studies [18]. Very recently, the optical measurements indicated that in $Ba(Fe_{1-x}P_x)_2As_2$ the electronic-nematic order emerges in a genuine phase transition [19].

A plausibility of such a scenario was also suggested by another recent studies of the in-plane thermoelectric anisotropy in the strain-detwinned $Ba(Fe_{1-x}Co_x)_2As_2$ iron-based superconductor [13]. Those results include data on the anisotropy of the Nernst effect ($\Delta v$) defined as a difference between the Nernst coefficient determined in the magnetic field parallel to the crystallographic $c$ axis with the thermal gradient imposed either along the $b$ axis ($v_b$) or the $a$ axis ($v_a$). The striking feature is that $\Delta v$ in Ba(Fe$_{1-x}$Co$_x$)$_2$As$_2$ is not affected by the magnetic/structural transition at $T_{tr}$ and $\Delta v(T)$ develops smoothly down to the lowest temperatures changing almost linearly in the semi-logarithmic scale. The onset of the Nernst anisotropy occurs at a temperature significantly higher than $T_{tr}$, although the fact that the magnetic/structural transition in Ba(Fe$_{1-x}$Co$_x$)$_2$As$_2$ is identified as being second order [20] does not allow one to exclude that non-zero $\Delta v$ comes as a result of the critical nematic or magnetic fluctuations. The case of Ca(Fe$_{1-x}$Co$_x$)$_2$As$_2$ discussed here is different because the transition in this compound has been classified as being first-order (discontinuous) [21,22], thus no substantial contribution from fluctuations is expected.

**Experiment**

Single crystals of Ca(Fe$_{1-x}$Co$_x$)$_2$As$_2$ were grown using the Sn-flux method. The Ca, Fe, As, Co, and Sn elements in molar ratios of 1 : (2-$x$) : 2 : $x$ : 30 ($x$ = 0, 0.06 and 0.2 respectively for CaCo0 CaCo3 and CaCo7) were loaded into alumina crucibles and sealed in quartz ampules under vacuum. The ampules were heated slowly to 1050 C, kept at this temperature for several hours, and then cooled down slowly to 650 C at a rate of 2 C/h. Next, the liquid tin was decanted from the crucibles. The Sn residues on the crystals were removed via etching in diluted hydrochloric acid. The chemical composition of a Co-doped single crystal was determined by the energy dispersive x-ray analysis.

The Hall coefficient was measured in unstrained crystals in a magnetic field of $B$ = 12.5 T. Then a sample was mounted between two clamps made of phosphor bronze and subjected to a uniaxial pressure applied along its sides by a beryllium copper spring controlled with a stepper motor. For the resistivity measurements, the electrical contacts were placed at the corners of a sample and the orientations of the voltage and current leads were switched repetitively during the experiment. This allowed the electrical resistivities $\rho_a$ and $\rho_b$ to be determined using the Montgomery method [23]. The uniaxial pressure was increased step-by-step and measurements of the resistivity were repeated until a saturation of the anisotropy, indicating maximal detwinning, was achieved. The maximal pressure determined in this way was used during subsequent thermoelectric experiments.

The Nernst coefficient was measured along and across the strain direction in two separate runs with the magnetic field (parallel to *c*-axis) varied from -12.5 T to +12.5 T. The temperature difference along a sample was determined using two Cernox thermometers as well as a calibrated in magnetic field constantan – chromel thermocouple attached to the sample through a few millimetres long and 100 μm thick silver wires. Signal leads were made up of long pairs of 25 μm phosphor bronze wires. $v_a$ and $v_b$, which differ within several percent due to slightly different geometrical factors in two experimental configurations are matched by applying a multiplicative correction factor close to 1. More details about the experimental setup are given in Ref. 3.

**Results and Discussion**

Figure 1*a* presents electrical resistivity measurements performed along the long (*a*) and short (*b*) crystallographic axes in a series of the strain-detwinned Ca(Fe$_{1-x}$Co$_x$)$_2$As$_2$ single crystals with $x = 0$ (CaCo0), 0.03 (CaCo3), and 0.07 (CaCo7). In all samples the transition to the orthorhombic SDW state affect mostly the electrical transport along the *b* axis. The anisotropy of the resistivity disappears rather quickly in the tetragonal state as shown in Fig. 1*b*, where the normalized temperature dependences of the resistive anisotropy ($\Delta\rho = (\rho_b - \rho_a)/(\rho_b + \rho_a)$) form a sharp step in CaCo0 at 167 K, a two steps transition in CaCo3

at 148 K (where the structural and magnetic transitions are separated by about 10 K) and a rather smooth development of $\Delta\rho$ below 100 K in CaCo7. The Hall coefficient, which temperature dependences in the vicinity of the transitions are presented in Fig. 1*c*, exhibits an anomaly at the respective temperatures in all three samples, which is not exactly the case for the Nernst coefficient presented in Fig. 2*a*. While $v(T)$ in CaCo0 and CaCo3 exhibit a clear step-like change at the structural/magnetic transition (which might be related to the presence of the Dirac fermions in the SDW phase of the 122 iron based superconductors [24,25]) such an anomaly is absent in CaCo7. Figure 2*b*, presenting the temperature and doping evolution of the Nernst anisotropy calculated as $\Delta v = v_a - v_b$, indicates that $\Delta v$ in CaCo0 and CaCo3 changes sign below $T_{tr}$, whereas for all three samples $\Delta v$ is positive in the tetragonal state. This is different from Ba(Fe$_{1-x}$Co$_x$)$_2$As$_2$, where $(v_a - v_b)$ is negative for all samples being studied except of the most doped Ba(Fe$_{0.94}$Co$_{0.06}$)$_2$As$_2$, where $\Delta v$ is small but positive. Interestingly the onset of $\Delta v$ shown in Fig. 2*c* occurs at the temperatures significantly higher than each respective $T_{tr}$. Namely, they are: $T^* \approx 275$, 280 and 250 K for CaCo0, CaCo3, and CaCo7, respectively. Remarkably, these values placed on the *x-T* phase diagram shown in Fig. 3 closely correspond to the analogous temperatures determined for the Ba(Fe$_{1-x}$Co$_x$)$_2$As$_2$ series [13]. The phase diagrams for both compounds are normalised to match the range of existence of the SDW phase, ie. *T* values are divided by the $T_{SDW}$ for respective parent compounds (BaFe$_2$As$_2$ or CaFe$_2$As$_2$), whereas *x* values are divided by the respective $x_{SDW}$ that is the cobalt content at which SDW disappears in $T = 0$ K limit.

Perhaps the most important outcome of this work is the set of temperature dependences of the Nernst anisotropy shown in Fig. 4*a*. We notice that $\Delta v$ for Ca(Fe$_{1-x}$Co$_x$)$_2$As$_2$ has opposite sign to $\Delta v$ for Ba(Fe$_{1-x}$Co$_x$)$_2$As$_2$ (except for BaCo6) but it is not trivial to explain. The normal state Nernst coefficient is composed of two terms: $v = \left(\frac{\alpha_{xy}}{\sigma} - S\tan\theta\right)\frac{1}{B}$ (where $\alpha_{xy}$ is the off-diagonal element of the Peltier tensor, and $\theta$ is the Hall angle) and its sign depends on details of the electronic structure and scattering processes. The explanation of the sign change of the ($v_a$ - $v_b$) difference between Ca(Fe$_{1-x}$Co$_x$)$_2$As$_2$ and Ba(Fe$_{1-x}$Co$_x$)$_2$As$_2$, which is unlikely incidental, would require much more elaborated approach than one offered in the present work. Therefore, in Fig. 4*a* plotted is the absolute value of the Nernst anisotropy divided by *T* to account for the fact that the Nernst coefficient, being a measure of transverse entropy flow [26], inevitably decreases with decreasing temperature. At the structural/magnetic transition $|\Delta v|/T$ changes abruptly in CaCo0 and CaCo3, perhaps due to a potent contribution to the Nernst coefficient from the afore mentioned Dirac fermions or as a signature of the strong first order character of the transition [8]. In contrast, $|\Delta v|/T$ in CaCo7 seems to be unaffected by the formation of the orthorhombic/SDW state. Notably, the character of the $|\Delta v|/T$ temperature dependence is shared among all the samples studied, and behaviour of $|\Delta v|/T(T)$ is far from the $\frac{C}{T-T_c}$ form expected for the Curie-Weiss behaviour. The latter was suggested to be caused (with some deviations) by critical fluctuations of the nematic state in iron-based superconductors studied by means of elastoresistance measurements [12]. As a matter of fact, our data are much better fitted by the equation proposed by I. Haller to describe behaviour of the nematic order parameter *Q* in liquid crystals [27]: $Q = (1 - \frac{T}{T^*})^\beta$, where *T\** is a characteristic temperature somewhat lower than the actual temperature of the discontinuous nematic-isotropic transition. The actual fits: $\frac{|\Delta v|}{T} = \alpha(1 - \frac{T}{T^*})^\beta$ have only two free parameters (*α* and *β*) since *T\** values are taken as the temperatures of the Nernst anisotropy onset (Fig. 2*c*). Here *α* is the proportional constant and *β* is the pseudo-critical exponent. Some deviations from the fitting curves near *T\** could be a result of the empirical nature of the Haller's approximation equation, which is not expected to be accurate close to the transition temperature [28].

The pseudo-critical exponent *β* shown in Fig. 4*b* stays within the 1 - 4 range for (Ca,Ba)(Fe$_{1-x}$Co$_x$)$_2$As$_2$, which is about one order of magnitude higher in than values observed in liquid crystals [27,28]. According to A. Ranjkesh et al. the pseudo-critical exponent is supposed to rise with concentrations of impurities [28] and estimated $\beta$ in (Ba,Co)(Fe$_{1-x}$Co$_x$)$_2$As$_2$ seems to roughly obey this rule increasing with $x$ (since growing substitution of iron with cobalt leads inevitably to an increased disorder). Although, the data scattering does not allow us to draw a definitive conclusion.

Here we would like to notice that the data from an analogous experiment performed on YBa$_2$Cu$_3$O$_y$ [15], presented in Fig 4*c*, can also be very well fitted with the Haller equation (this time $T^*$ is taken as a free parameter). Such a universal behaviour of the $|\Delta v|/T$ temperature dependence along with the fact that $|\Delta v|/T$ is unaffected by the structural/magnetic transition in CaCo7 as well as in Ba(Fe$_{1-x}$Co$_x$)$_2$As$_2$ with $x = 0.02$ and $0.04$ [13], suggests the anisotropy of the Nernst coefficient might be a proxy of the electronic-nematic order parameter. This idea was already considered by R. Daou et al. [16], which were inspired by the analogous proposition by R.A. Borzi et al. [29] to define the phenomenological nematic order parameter in Sr$_3$Ru$_2$O$_7$ as the in-plane anisotropy of the electrical resistivity. A remaining question is why we observe in the thermoelectrical studies the broken rotational symmetry, which is inaccessible to electrical measurements and vice versa – that is, why the anomaly at the transition clearly visible in $\Delta\rho(T)$ is absent in $\Delta v(T)$? While we cannot provide a firm answer, there are possible scenarios that could explain such an intriguing behaviour. First, there are well known examples of electronic systems with decoupled charge and entropy fluxes. One of them is the superconducting state where Cooper pairs conduct charge without dissipation but do not contribute in entropy transport. Another are small angle electron-phonon scattering processes, which effectively disturb entropy transport with only little effect on the charge transfer. Remarkably, the latter play an important role in strongly correlated electron system, where charge transport could be described as the flow of a hydrodynamic electron fluid [30]. In such a case the energy dissipation is dominated by momentum conserving small angle electron-phonon scattering [31]. Therefore, if charge carriers in the electronic-nematic state suffer from the anisotropic, but small-angle scattering, it will be seen in the anisotropy of the Nernst effect, but be insignificant for the electrical resistivity measurements. A similar phenomenon was recently observed in $YBa_2Cu_3O_y$ and attributed to the emergence of the short-range charge density wave modulations [15]. On the other hand, a possible reason why the anomaly at the magnetic/structural transition is absent in the $|\Delta \nu|/T$ temperature dependence was proposed in Ref. 3. Namely, the large angle elastic scattering might cancel out for the Nernst anisotropy which consists of contributions from both *a*- and *b*-direction transport. In fact, $\Delta \nu$ can be expressed as a subtraction of a two Sondheimer cancellations [32], one for *a*-direction, another one for *b*-direction [13], and in metals the resulting difference might be insignificant.

**Conclusion**

In summary, we show that in the phase diagram of iron-based superconductors, in a way similar to their copper-based counterparts, there is room for another electronically driven ordered state. Using the Nernst anisotropy as a proxy of the electronic-nematic order parameter we indicate the emergence of the high temperature true nematic phase. Despite we are not able to pinpoint its possible origin [8], we stress the role of the elastic and inelastic scattering in the process. The electronic-nematic phase that arises at high temperature and can develop smoothly despite the occurrence of the magnetic/structural transition. The present form of the phase diagram does not exclude that the electronic-nematic phase ceases when the superconducting critical temperature reaches its maximum, which would once again be a common behaviour for iron-based and copper-based supercoducters [33]. Perhaps it might even suggest an intimate relation between the electronic nematicity and superconductivity [34].

**Acknowledgments**

The authors wish to thank J.R. Cooper, G.G. Ihas, K. Rogacki and T. Kopeć for valuable discussions.

This work was supported financially by the National Science Centre (Poland) under the research Grant No. 2014/15/B/ST3/00357.

**M.Matusiak ORCID iD**: 0000-0003-4480-9373

**Figures**

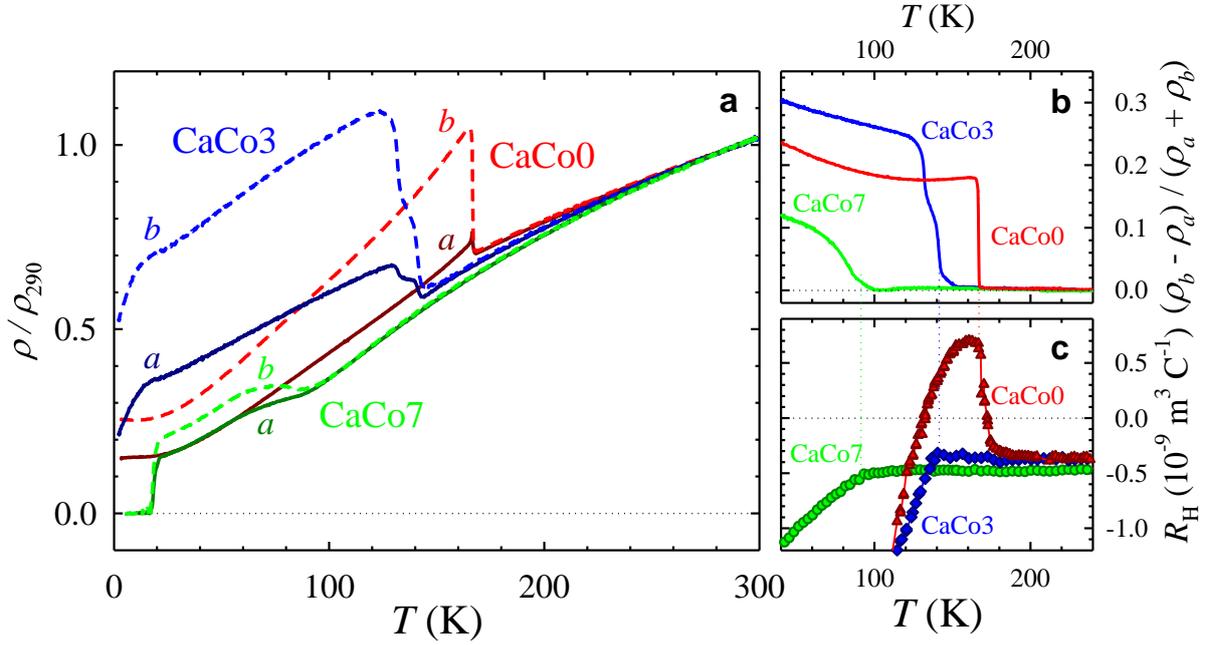

**Figure 1.**

(Color online) **In-plane electrical transport in the $Ca(Fe_{1-x}Co_x)_2As_2$ series. a,** Temperature dependences of the in-plane electrical resistivity (normalised to the high temperature, $T$ = 290 K, value) for the detwinned single crystals. There is a clear difference between $\rho$ measured along the long ($a$) and short ($b$) crystallographic axis in the orthorhombic/SDW state. **b,** Temperature dependences of the normalised resistivity anisotropy $(\rho_b - \rho_a)/(\rho_a + \rho_b)$ showing no sign of a significant difference between $\rho_a$ and $\rho_b$ high above the structural transition. **c,** Anomalies in the temperature dependences of the Hall coefficient occurring at the structural/magnetic transition.

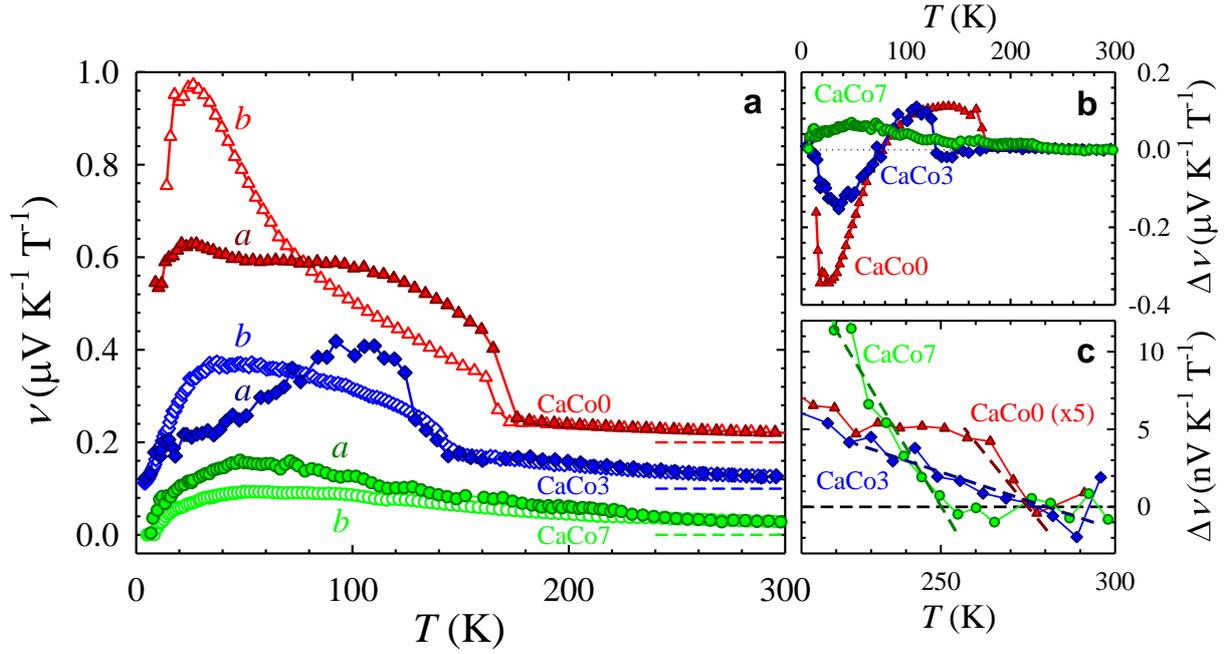

**Figure 2.**

(Color online) **In-plane magneto-thermoelectric response in the Ca(Fe$_{1-x}$Co$_x$)$_2$As$_2$ series. a**, Temperature dependences of the Nernst coefficient for the heat current being imposed along either *a* or *b* orthorhombic axis. CaCo3 and CaCo0 plots are shifted vertically by 1 x 10$^{-10}$ and 2 x 10$^{-10}$ V K$^{-1}$ T$^{-1}$, respectively, for the sake of clarity with dashed lines denoting the corresponding ν = 0 V K$^{-1}$ T$^{-1}$ levels. **b**, Temperature dependence of the Nernst coefficient anisotropy defined as Δν = ν$_a$ − ν$_b$. **c**, High temperature onset of the Nernst anisotropy.

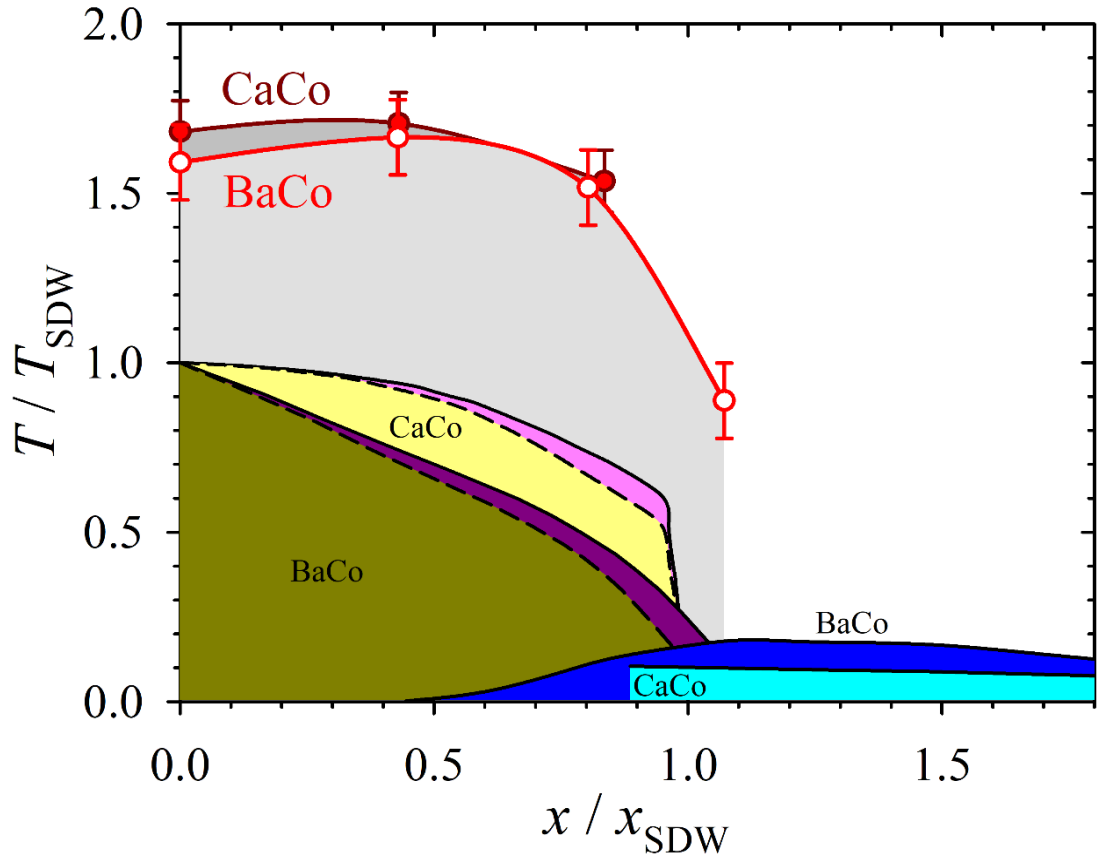

**Figure 3.**

(Color online) **Schematic *x-T* phase diagram of Ca(Fe$_{1-x}$Co$_x$)$_2$As$_2$ [35] and Ba(Fe$_{1-x}$Co$_x$)$_2$As$_2$ [36] with marked positions of the onset of the Nernst anisotropy.** Results for Ca(Fe$_{1-x}$Co$_x$)$_2$As$_2$ are marked with full red points and for Ba(Fe$_{1-x}$Co$_x$)$_2$As$_2$ with hollow red points. Yellow and dark yellow regions in the phase diagram denote spin density wave phase; pink and dark pink -- orthorhombic, nonmagnetic phase; blue and dark blue -- superconducting phase.

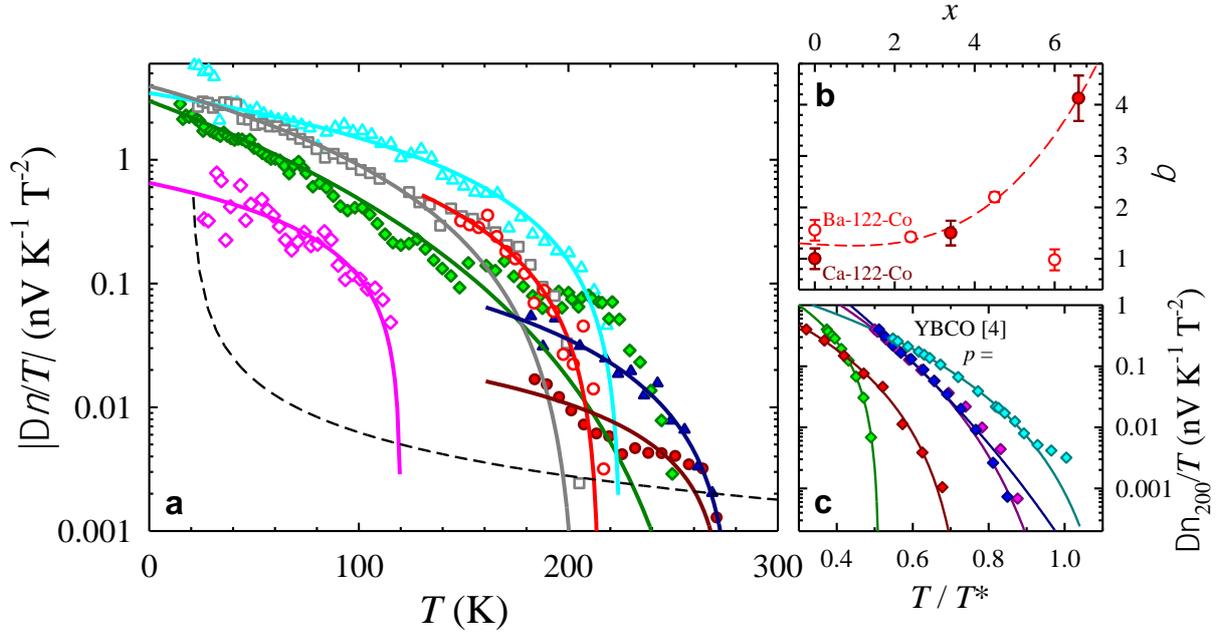

**Figure 4.**

(Color online) **Anisotropy of the Nernst coefficient as a measure of the electronic nematic order parameter in iron- and copper-based superconductors. a**, Temperature dependences of the absolute value of $\Delta\nu/T$ for the Ca(Fe$_{1-x}$Co$_x$)$_2$As$_2$ (solid symbols) and Ba(Fe$_{1-x}$Co$_x$)$_2$As$_2$ [13] (hollow symbols) series, where "(+)" and "(-)" marks denote the positive and negative sign of $\Delta\nu/T$, respectively. Solid lines are fits to the temperature dependence of the nematic order parameters considered in liquid crystals [27], whereas dashed line is an exemplary Curie – Weiss type dependence $\Delta\nu/T = \frac{5*10^{-10}}{T-20}$. **b**, Changes of the $\beta$ parameter with doping that are supposedly reflecting the increasing level of disorder. **c**, Nernst anisotropy in YBa$_2$Cu$_3$O$_y$, where $\Delta\nu_{200} = \Delta\nu(T) - \Delta\nu(200\ K)$ and the temperature axis is also normalized [15]. Solid lines give fits for the nematic order parameter as mentioned in the text.

**References**

[1] A.W. Sleight, *Chemistry of High-Temperature Superconductors*, Science **242**, 1519 (1988), DOI: 10.1126/science.242.4885.1519;

[2] H. Ding, T. Yokoya, J. C. Campuzano, T. Takahashi, M. Randeria, M. R. Norman, T. Mochiku, K. Kadowaki & J. Giapintzakis, *Spectroscopic evidence for a pseudogap in the normal state of underdoped high-Tc superconductors*, Nature **382**, 51 (1996), https://doi.org/10.1038/382051a0;

[3] D. Haug, V. Hinkov, Y. Sidis, P. Bourges, N.B. Christensen, A. Ivanov, T. Keller, C.T. Lin and B. Keimer, *Neutron scattering study of the magnetic phase diagram of underdoped $YBa_2Cu_3O_{6+x}$*, New Journal of Physics **12**, 105006 (2010), doi:10.1088/1367-2630/12/10/105006;

[4] Tao Wu, Hadrien Mayaffre, Steffen Krämer, Mladen Horvatić, Claude Berthier, W. N. Hardy, Ruixing Liang, D. A. Bonn & Marc-Henri Julien, *Magnetic-field-induced charge-stripe order in the high-temperature superconductor $YBa_2Cu_3O_y$*, Nature **477**, 191 (2011), https://doi.org/10.1038/nature10345;

[5] Francis Laliberté, Mehdi Frachet, Siham Benhabib, Benjamin Borgnic, Toshinao Loew, Juan Porras, Mathieu Le Tacon, Bernhard Keimer, Steffen Wiedmann, Cyril Proust & David LeBoeuf, *High field charge order across the phase diagram of $YBa_2Cu_3O_y$*, npj Quantum Materials **3**, 11 (2018), https://doi.org/10.1038/s41535-018-0084-5;

[6] Eduardo Fradkin, Steven A. Kivelson, Michael J. Lawler, James P. Eisenstein, Andrew P. Mackenzie, *Nematic Fermi Fluids in Condensed Matter Physics*, Annual Review of Condensed Matter Physics **1**, 153 (2010), https://doi.org/10.1146/annurev-conmatphys-070909-103925;

[7] J.Wu, A.T.Bollinger, X.He, I.Božović, Pervasive electronic nematicity in a cuprate superconductor, Physica C: Superconductivity and its Applications **549**, 95 (2018), https://doi.org/10.1016/j.physc.2018.02.056;

[8] R. M. Fernandes, A. V. Chubukov & J. Schmalian, *What drives nematic order in iron-based superconductors?*, Nature Physics **10**, 97 (2014), https://doi.org/10.1038/nphys2877;

[9] S. Kasahara, H.J. Shi, K. Hashimoto, S. Tonegawa, Y. Mizukami, T. Shibauchi, K. Sugimoto, T. Fukuda, T. Terashima, Andriy H. Nevidomskyy & Y. Matsuda, *Electronic nematicity above the structural and superconducting transition in $BaFe_2(As_{1-x}P_x)_2$*, Nature 486, 382 (2012), https://doi.org/10.1038/nature11178;

[10] S. Avci, O. Chmaissem, J.M. Allred, S. Rosenkranz, I. Eremin, A.V. Chubukov, D.E. Bugaris, D.Y. Chung, M.G. Kanatzidis, J.-P Castellan, J.A. Schlueter, H. Claus, D.D. Khalyavin, P. Manuel, A. Daoud-Aladine & R. Osborn, Magnetically driven suppression of nematic order in an iron-based superconductor, Nat. Commun. **5**, 3845 (2014), https://doi.org/10.1038/ncomms4845;

[11] E.C. Blomberg, M.A. Tanatar, R.M. Fernandes, I.I. Mazin, Bing Shen, Hai-Hu Wen, M.D. Johannes, J. Schmalian & R. Prozorov, *Sign-reversal of the in-plane resistivity anisotropy in hole-doped iron pnictides*, Nat. Commun. **4**, 1914 (2013), doi: 10.1038/ncomms2933 (2013);

[12] Hsueh-Hui Kuo, Jiun-Haw Chu, Johanna C. Palmstrom, Steven A. Kivelson, Ian R. Fisher, *Ubiquitous signatures of nematic quantum criticality in optimally doped Fe-based superconductors*, Science 352, 958 (2016), DOI: 10.1126/science.aab0103;

[13] Marcin Matusiak, Krzysztof Rogacki, and Thomas Wolf, *Thermoelectric anisotropy in the iron-based superconductor $Ba(Fe_{1-x}Co_x)_2As_2$*, Phys. Rev. B **97**, 220501(R) (2018), https://doi.org/10.1103/PhysRevB.97.220501;

[14] Marcin Matusiak, Michał Babij, and Thomas Wolf, *Anisotropy of the Seebeck and Nernst coefficients in parent compounds of the iron-based superconductors*, Phys. Rev. B 97, 100506(R), https://doi.org/10.1103/PhysRevB.97.100506;

[15] O. Cyr-Choinière, G. Grissonnanche, S. Badoux, J. Day, D. A. Bonn, W. N. Hardy, R. Liang, N. Doiron-Leyraud, and Louis Taillefer, *Two types of nematicity in the phase diagram of the cuprate superconductor $YBa_2Cu_3O_y$*, Phys. Rev. B **92**, 224502 (2015), https://doi.org/10.1103/PhysRevB.92.224502;

[16] R. Daou, J. Chang, David LeBoeuf, Olivier Cyr-Choinière, Francis Laliberté, Nicolas Doiron-Leyraud, B. J. Ramshaw, Ruixing Liang, D. A. Bonn, W. N. Hardy & Louis Taillefer, *Broken rotational*

*symmetry in the pseudogap phase of a high-Tc superconductor*, Nature **463**, 519 (2010), https://doi.org/10.1038/nature08716;

[17] M.A. Tanatar, E.C. Blomberg, A. Kreyssig, M.G. Kim, N. Ni, A. Thaler, S.L. Bud'ko, P.C. Canfield, A.I. Goldman, I.I. Mazin, and R. Prozorov, *Uniaxial-strain mechanical detwinning of $CaFe_2As_2$ and $BaFe_2As_2$ crystals: Optical and transport study*, Phys. Rev. B **81**, 184508 (2010), DOI: 10.1103/PhysRevB.81.184508;

[18] H. Z. Arham, C. R. Hunt, W. K. Park, J. Gillett, S. D. Das, S. E. Sebastian, Z. J. Xu, J. S. Wen, Z. W. Lin, Q. Li, G. Gu, A. Thaler, S. L. Budko, P. C. Canfield, and L. H. Greene, *Gap-like feature in the normal state of $X(Fe_{1-x}Co_x)_2As_2$, X = Ba; Sr and $Fe_{1+y}Te$ revealed by Point Contact Spectroscopy*, J. Phys. Conf. Ser. **400**, 022001 (2012), doi:10.1088/1742-6596/400/2/022001;

[19] Eric Thewalt, Ian M. Hayes, James P. Hinton, Arielle Little, Shreyas Patankar, Liang Wu, Toni Helm, Camelia V. Stan, Nobumichi Tamura, James G. Analytis, and Joseph Orenstein, *Imaging Anomalous Nematic Order and Strain in Optimally Doped $BaFe(As,P)_2$*, Phys. Rev. Lett. **121**, 027001 (2018), DOI:10.1103/PhysRevLett.121.027001;

[20] D.K. Pratt, W. Tian, A. Kreyssig, J.L. Zarestky, S. Nandi, N. Ni, S.L. Bud'ko, P.C. Canfield, A.I. Goldman, and R.J. McQueeney, *Coexistence of Competing Antiferromagnetic and Superconducting Phases in the Underdoped $Ba(Fe_{0.953}Co_{0.047})_2As_2$ Compound Using X-ray and Neutron Scattering Techniques*, Phys. Rev. Lett. **103**, 087001 (2009), https://doi.org/10.1103/PhysRevLett.103.087001;

[21] Sergey L. Bud'ko, Sheng Ran, and Paul C. Canfield, *Thermal expansion of $CaFe_2As_2$: Effect of cobalt doping and postgrowth thermal treatment*, Phys. Rev. B 88, 064513 (2013), https://doi.org/10.1103/PhysRevB.88.064513;

[22] Antón Fente, Alexandre Correa-Orellana, Anna E. Böhmer, Andreas Kreyssig, S. Ran, Sergey L. Bud'ko, Paul C. Canfield, Federico J. Mompean, Mar García-Hernández, Carmen Munuera, Isabel Guillamón, and Hermann Suderow, *Direct visualization of phase separation between superconducting and nematic domains in Co-doped $CaFe_2As_2$ close to a first-order phase transition*, Phys. Rev. B **97**, 014505 (2018), https://doi.org/10.1103/PhysRevB.97.014505;

[23] H.C. Montgomery, *Method for Measuring Electrical Resistivity of Anisotropic Materials*, Journal of Applied Physics **42**, 2971 (1971), DOI: 10.1063/1.1660656.

[24] T. Morinari, E. Kaneshita, and T. Tohyama, *Topological and Transport Properties of Dirac Fermions in an Antiferromagnetic Metallic Phase of Iron-Based Superconductors*, Phys. Rev. Lett. 105, 037203 (2010), https://doi.org/10.1103/PhysRevLett.105.037203;

[25] Z.-G. Chen, L. Wang, Y. Song, X. Lu, H. Luo, C. Zhang, P. Dai, Z. Yin, K. Haule, and G. Kotliar, *Two-Dimensional Massless Dirac Fermions in Antiferromagnetic $AFe_2As_2$ (A=Ba,Sr)*, Phys. Rev. Lett. 119, 096401 (2017), https://doi.org/10.1103/PhysRevLett.119.096401;

[26] Kamran Behnia and Hervé Aubin, *Nernst effect in metals and superconductors: a review of concepts and experiments*, Rep. Prog. Phys. **79**, 046502 (2016), doi:10.1088/0034-4885/79/4/046502;

[27] Ivan Haller, *Thermodynamic and static properties of liquid crystals*, Solid State Chem. **10**, 103 (1975), https://doi.org/10.1016/0079-6786(75)90008-4;

[28] Amid Ranjkesh, Matej Cvetko, Jun-Chan Choi & Hak-Rin Kim, *Phase and structural order in mixture of nematic liquid crystals and anisotropic nanoparticles*, Phase Transitions **90**, 423 (2017), DOI: 10.1080/01411594.2016.1212197;

[29] R.A. Borzi, S. A. Grigera, J. Farrell, R.S. Perry, S.J.S. Lister, S.L. Lee, D.A. Tennant, Y. Maeno, A. P. Mackenzie, *Formation of a Nematic Fluid at High Fields in $Sr_3Ru_2O_7$*, Science **315**, 214 (2007), DOI: 10.1126/science.1134796;

[30] J. Gooth, F. Menges, N. Kumar, V. Süβ, C. Shekhar, Y. Sun, U. Drechsler, R. Zierold, C. Felser & B. Gotsmann, *Thermal and electrical signatures of a hydrodynamic electron fluid in tungsten diphosphide*, Nat. Commun. **9**, 4093 (2018); 9: 4093, https://doi.org/10.1038/s41467-018-06688-y;

[31] R.N Gurzhi, *Hydrodynamic effects in solids at low temperature*, Sov. Phys. Usp. **11**, 255 (1968), http://dx.doi.org/10.1070/PU1968v011n02ABEH003815;

[32] E.H. Sondheimer, *The theory of the galvanomagnetic and thermomagnetic effects in metals*, Proc. R. Soc. London, Ser. A **193**, 484 (1948), DOI: 10.1098/rspa.1948.0058;

[33] Cyril Proust, Louis Taillefer, *The remarkable underlying ground states of cuprate superconductors*, Annual Review Condensed Matter Physics **10**, 409 (2019), DOI: 10.1146/annurev-conmatphys-031218-013210;

[34] S. Lederer, Y. Schattner, E. Berg, and S.A. Kivelson, *Enhancement of Superconductivity near a Nematic Quantum Critical Point*, Phys. Rev. Lett. **114**, 097001 (2015), https://doi.org/10.1103/PhysRevLett.114.097001;

[35] L. Harnagea, S. Singh, G. Friemel, N. Leps, D. Bombor, M. Abdel-Hafiez, A. U. B. Wolter, C. Hess, R. Klingeler, G. Behr, S. Wurmehl, and B. Büchner, *Phase diagram of the iron arsenide superconductors $Ca(Fe_{1-x}Co_x)_2As_2 (0 \leqslant x \leqslant 0.2)$*, Phys. Rev. B **83**, 094523 (2011), https://doi.org/10.1103/PhysRevB.83.094523;

[36] Jiun-Haw Chu, James G. Analytis, Chris Kucharczyk, and Ian R. Fisher, *Determination of the phase diagram of the electron-doped superconductor $Ba(Fe_{1-x}Co_x)_2As_2$*, Phys. Rev. B **79**, 014506 (2009), https://doi.org/10.1103/PhysRevB.79.014506.